\documentstyle[12pt,epsfig]{article}

\newskip\humongous \humongous=0pt plus 1000pt minus 1000pt

\newif\ifdtup

\def\pr#1{#1^\prime}

\def\beq{\begin{equation}}
\def\eeq{\end{equation}}
\def\eq{\beq\eeq}
\def\beqn{\begin{eqnarray}}
\def\eeqn{\end{eqnarray}}
\relax

\def\dotx{\dotx{\dot\overline{x}}}

\relax

\catcode`\@=11

\@addtoreset{equation}{section}
\def\theequation{\thesection\arabic{equation}}

\def\@normalsize{\@setsize\normalsize{15pt}\xiipt\@xiipt
\abovedisplayskip 14pt plus3pt minus3pt%
\belowdisplayskip \abovedisplayskip
\abovedisplayshortskip \z@ plus3pt%
\belowdisplayshortskip 7pt plus3.5pt minus0pt}

\def\small{\@setsize\small{13.6pt}\xipt\@xipt
\abovedisplayskip 13pt plus3pt minus3pt%
\belowdisplayskip \abovedisplayskip
\abovedisplayshortskip \z@ plus3pt%
\belowdisplayshortskip 7pt plus3.5pt minus0pt
\def\@listi{\parsep 4.5pt plus 2pt minus 1pt
     \itemsep \parsep
     \topsep 9pt plus 3pt minus 3pt}}

\@twosidetrue

\relax

\catcode`@=12

\catcode`\@=11

\def\section{\@startsection{section}{1}{\z@}{3.5ex plus 1ex minus
   .2ex}{2.3ex plus .2ex}{\large\bf}}

\def\thesection{\arabic{section}.}

\def\appendix{\setcounter{section}{0}
 \def\thesection{APPENDIX \Alph{section}:}
 \def\theequation{\Alph{section}.\arabic{equation}}}

\def\ps@headings{\def\@oddfoot{}\def\@evenfoot{}
\def\@oddhead{\hbox{}\hfill
 \makebox[.5\textwidth]{\raggedright\ignorespaces --\thepage{}--
 \hfill {}}}  
\def\@evenhead{\@oddhead}
\def\subsectionmark##1{\markboth{##1}{}}
}

\ps@headings

\catcode`\@=12

\def\figcap{\section*{Figure Captions\markboth
 {FIGURECAPTIONS}{FIGURECAPTIONS}}\list
 {Fig. \arabic{enumi}:\hfill}{\settowidth\labelwidth{Fig. 999:}
 \leftmargin\labelwidth
 \advance\leftmargin\labelsep\usecounter{enumi}}}
 \relax
\def\tablecap{\section*{Table Captions\markboth
 {TABLECAPTIONS}{TABLECAPTIONS}}\list
 {Table \arabic{enumi}:\hfill}{\settowidth\labelwidth{Table 999:}
 \leftmargin\labelwidth
 \advance\leftmargin\labelsep\usecounter{enumi}}}
 \relax
\def\reflist{\section*{References\markboth
 {REFLIST}{REFLIST}}\list
 {[\arabic{enumi}]\hfill}{\settowidth\labelwidth{[999]}
 \leftmargin\labelwidth
 \advance\leftmargin\labelsep\usecounter{enumi}}}
 \relax

\catcode`\@=11

\def\ps@headings{\def\@oddfoot{}\def\@evenfoot{}
\def\@oddhead{\hbox{}\hfill
 \makebox[.5\textwidth]{\raggedright\ignorespaces --\thepage{}--
 \hfill {}}}    
\def\@evenhead{\@oddhead}
\def\subsectionmark##1{\markboth{##1}{}}
}

\ps@headings

\relax

\relax
\def\pl#1#2#3{{\it Phys. Lett. }{\bf #1}(19#2)#3}
\def\zp#1#2#3{{\it Z. Phys. }{\bf #1}(19#2)#3}
\def\prl#1#2#3{{\it Phys. Rev. Lett. }{\bf #1}(19#2)#3}

\def\pr#1#2#3{{\it Phys. Rev. }{\bf #1}(19#2)#3}
\def\np#1#2#3{{\it Nucl. Phys. }{\bf #1}(19#2)#3}

\relax

\begin{document}
\renewcommand\floatpagefraction{.95}
\renewcommand\topfraction{.95}
\renewcommand\bottomfraction{.95}
\renewcommand\textfraction{.05}
\setcounter{topnumber}{5}          
\setcounter{bottomnumber}{5}       
\setcounter{totalnumber}{5}        
\setcounter{dbltopnumber}{2}       
\newcommand\ssmallfig{3cm}
\newcommand\smallfig{4cm}
\newcommand\mediumfig{5cm}
\newcommand\bigfig{6cm}
\newcommand\bbigfig{9cm}
\def\theequation{\arabic{equation}}
\newcommand\as{\alpha_S}
\newcommand\aem{\alpha_{\rm em}}
\newcommand\refq[1]{$^{[#1]}$}
\newcommand\avr[1]{\left\langle #1 \right\rangle}
\newcommand\lambdamsb{
\Lambda_5^{\rm \scriptscriptstyle \overline{MS}}
}
\newcommand\qqb{{q\overline{q}}}
\newcommand\asb{\as^{(b)}}
\newcommand\qb{\overline{q}}
\newcommand\sigqq{\sigma_{q\overline{q}}}
\newcommand\siggg{\sigma_{gg}}
\newcommand\MSB{{\rm \overline{MS}}}
\def \eq {e_{\scriptscriptstyle Q}}
\def \muf {\mu_{\scriptscriptstyle F}}
\def \mur {\mu_{\scriptscriptstyle R}}
\def \muo {\mbox{$\mu_0$}}
\def \pt  {\mbox{$p_{\rm T}$}}
\def \ptg {\mbox{$p_{\rm T}^{q\overline{q}}$}}
\def \xf  {\mbox{$x_{\rm F}$}}
\def \dphi{\mbox{$\Delta\phi$}}
\def \dy  {\mbox{$\Delta y$}}
\def \pim {\mbox{$\pi^-$}}
\def \epem {\mbox{$e^+e^-$}}
\def \mc   {\mbox{$m_c$}}
\def \mb   {\mbox{$m_b$}}
\def \mqq   {\mbox{$M_{q\overline{q}}$}}
\def \tot   {{\rm tot}}
\newcommand\qq{{\scriptscriptstyle Q\overline{Q}}}
\newcommand\cm{{\scriptscriptstyle CM}}
\setcounter{topnumber}{10}
\setcounter{bottomnumber}{10}
\renewcommand\topfraction{0}
\renewcommand\textfraction{0}
\renewcommand\bottomfraction{1}
\newsavebox\tmpfig
\newcommand\settmpfig[1]{\sbox{\tmpfig}{\mbox{\cite{#1}}}}
%
\begin{titlepage}
\nopagebreak
\vspace*{-1in}
{\leftskip 11cm
\normalsize
\noindent
\newline
CERN-TH/95-256\newline IFUM 519/FT

}
\vskip 1.5cm
\begin{center}
{\large \bf \sc Heavy Flavour Production}
\vskip 1cm
{\bf Paolo Nason\footnotemark}\\
\footnotetext{On leave of absence from INFN, Sezione di Milano, Milan, Italy.}
{CERN TH-Division, CH-1211 Geneva 23, Switzerland}
\vskip 0.3cm
{\bf Stefano Frixione}\\
{Dip. di Fisica, Universit\`a di Genova, and INFN, Sezione di Genova,
Genoa, Italy}\\
\vskip .3cm
{\bf Giovanni Ridolfi\footnotemark}\\
\footnotetext{On leave of absence from INFN, Sezione di Genova, Genoa, Italy.}
{CERN TH-Division, CH-1211 Geneva 23, Switzerland}
\end{center}
\vskip 1cm
\nopagebreak
\begin{abstract}
{\small
We review the status of heavy flavour production in QCD.
Comparison of experimental and theoretical results for top
and bottom production are given.
Selected topics in charm production are also discussed.

}
\end{abstract}
\vskip 2.5cm
\begin{center}
Invited talk given at the

{\it XV International conference ``Physics in Collisions''}

Cracow, Poland, June 8-10, 1995.

\end{center}
\vfill
\noindent
CERN-TH/95-256 \newline IFUM 519/FT \newline September 1995    \hfill
\end{titlepage}

Although the first next-to-leading-order calculations of heavy flavour
production were performed more than five years
ago \cite{NDE} \cite{NDE2},
progress in this field is constantly being made. The work
of ref.~\cite{Neerven} has confirmed the results of ref.~\cite{NDE}.
A calculation of the next-to-leading cross section for the
photoproduction of heavy quarks has been given in ref.~\cite{EllisNason},
and has been confirmed by ref.~\cite{SmithNeerven}.
The computation of the radiative corrections to the electroproduction
of heavy quarks, via an off-shell photon, was presented in
ref.~\cite{LaenenPh}.
A method that accounts for the correlation of heavy
quarks at next-to-leading order was developed in ref.~\cite{MNR}
for heavy quark hadroproduction.
An application of this
calculation to fixed-target production of heavy quarks is given
in ref.~\cite{MNRFT}.
The method was extended to the photoproduction of heavy
quarks in ref.~\cite{FMNR}.

\begin{figure}[htb]
  \begin{center}
     \mbox{\epsfig{file=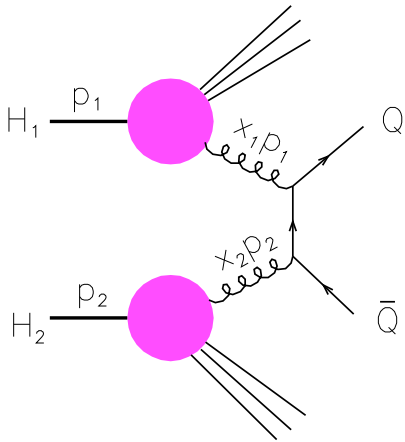,height=\bigfig}}
  \end{center}
\caption{\label{hvqproc}
The heavy quark production process in perturbative QCD.}
\end{figure}

In the following I will describe some recent experimental and theoretical
progress in top, bottom and charm production. I briefly remind the reader
of the QCD mechanism for heavy quark production.
The hadroproduction process is depicted in fig.~\ref{hvqproc}.
The high energy colliding hadrons can be viewed as
a broad band beam of partons (quarks and gluons), which collide and
fuse to produce the heavy quark pair. Thus the cross section is given by
the formula
\beq
\sigma_{Q\bar{Q}}=\int dx_1\,dx_2\; f^{H_1}_i\;(x_1,\mu)
f^{H_2}_k\;(x_2,\mu)\; \hat\sigma_{ij}(x_1\,p_1,\,x_2\,p_2,m_Q,\mu,\as)
\eeq
where $\hat\sigma$ is the short--distance cross section, which is calculable
order by order in perturbation theory. In fig.~\ref{hvqproc} only one
lowest order diagram is represented. Diagrams like the one of
fig.~\ref{hvqproc}, after squaring the amplitude
to obtain a cross section, give rise to a contribution of order $\as^2$.
Thus, the leading process is of order $\as^2$, and the next--to--leading
one is $\as^3$. The coupling constant $\as$ is evaluated
at a scale of the order of the mass of the heavy flavour. Thus, for charm
the coupling constant $\as$ is roughly 0.3 to 0.5, for bottom is
around 0.2, and for top is 0.1. We expect therefore that theoretical
predictions should be very reliable for top, less for bottom, and even less
for charm. In fact, next--to--leading corrections are around 30--40\%
for top, 100\% for bottom, and even larger and less controllable for charm.
It is clear therefore that for top we expect to be able to
predict the cross section rather well. For bottom, and expecially for
charm, we can expect deviations from theoretical predictions due
to higher order and non--perturbative effects. In these cases,
we also must keep in mind whether certain effects have a simple
explanation in terms of some non--perturbative effects, and whether
one can model the cross section with the perturbative result supplemented
with some model for the non--perturbative effects.

I will first deal with the issue of comparing the measured top cross section
to theoretical expectations. I will then discuss the status of bottom
production. Charm production physics is a rather
wide and complex field, mostly because of the fact that non-perturbative
effects in charm production
do play an important role. Thus, there are several experimental issues
of a certain interest, like the $A$ dependence, and the leading particle
effect. Here I will focus upon what I think are the most important
discrepancies between perturbation theory predictions and experimental
results. I will show that certain distributions, like the transverse
momentum of charmed mesons, the transverse momentum of the pair,
the azimuthal distance of the pair, and the invariant mass of the pair,
are somewhat in contrast with each other, and that they may give
an indication of how the comparison of theory and data should be performed.

\section{Top production}
\begin{figure}[htb]
\begin{center}\mbox{\epsfig{file=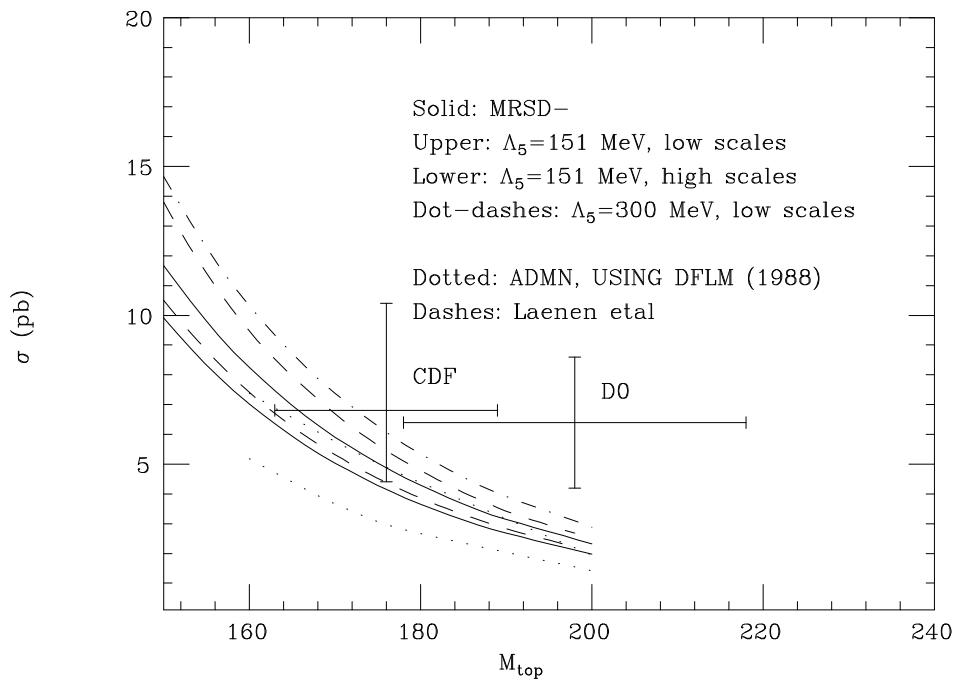,height=\bbigfig}}\end{center}
\caption{\label{topxsec}
Cross sections for top production: various calculations versus
the CDF and D0 measurements.
}
\end{figure}

For top hadroproduction, it was found that radiative
corrections are generally well under control.
This allows us to make predictions for top cross sections with a
relatively small error.
The recent measurements of top cross section at the Tevatron
\cite{CDFtop},\cite{D0top} have turned out to be in remarkable agreement
with the theoretical predictions. This is illustrated in fig.~\ref{topxsec}.
In the figure we have reported the result of the old calculation
of ref.~\cite{AltarelliTop}, using the DFLM structure functions,
a next--to--leading calculation using the MRSD-
set, and the calculation of ref.~\cite{LaenenTop} which attempts to
resummation of Sudakov threshold effects. Other calculations
are reported in ref~\cite{EllisTop} \cite{BergerTop},
where in the latter an alternative approach to the
inclusion of threshold effects is attempted.
When estimating the cross section, one should not forget that the uncertainty
on $\Lambda_5$, which defines the strenght of the QCD coupling constant,
is larger than what is usually assumed in structure function fits.
In fact, all LEP determinations \cite{Schmelling} tend to give a larger value.
Therefore, for comparison, we also show the computed
cross section in case one uses a value of $\Lambda_5$ more
compatible with LEP determination. The corresponding increase of the
cross section is presumably an overestimate, since a larger value
of $\Lambda_5$ implies stronger evolution of the parton densities,
which soften more rapidly as the momentum scale increases.
The figure shows that only minor changes in the
theoretical predictions have taken place in the last few years.
We can expect that in spite of the fact that more subleading contributions
may be added in the future to the top strong cross section, they
will not change substantially the pure next--to--leading result.
\section{Bottom production}
\begin{figure}[htb]
\begin{center}\mbox{\epsfig{file=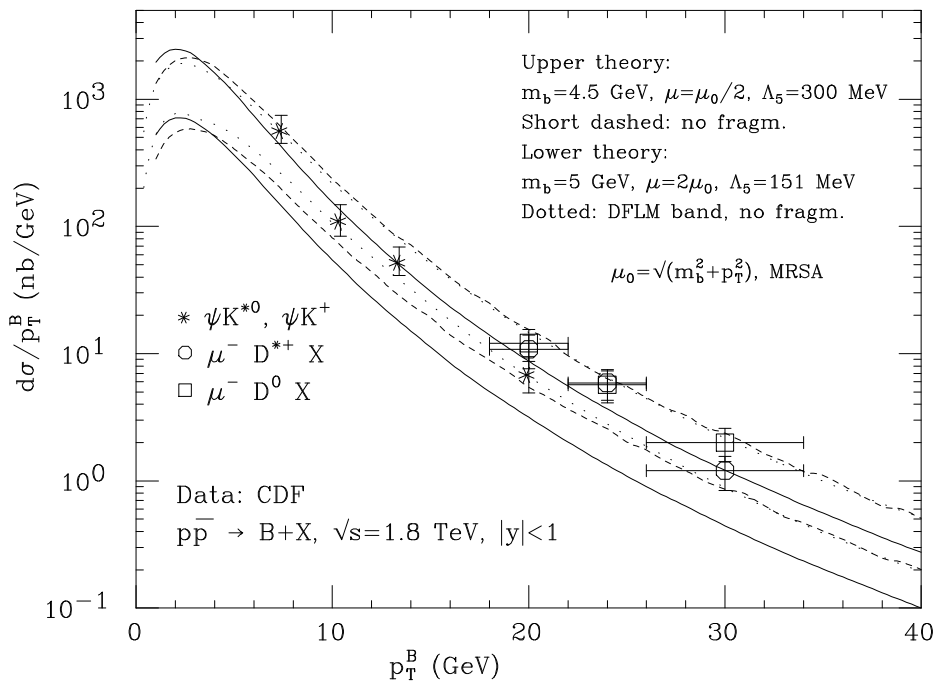,height=\bbigfig}}\end{center}
\caption[]{\label{cdfbpt}
Transverse momentum spectrum of B mesons at Tevatron energies,
as measured by the CDF collaboration. The main prediction band
(solid lines) is also shown without the inclusion of fragmentation
effects. The dotted lines are the old prediction of ref.~\box\tmpfig,
using the DFLM structure functions sets.
}
\end{figure}

\begin{figure}[htb]
\begin{center}\mbox{\epsfig{file=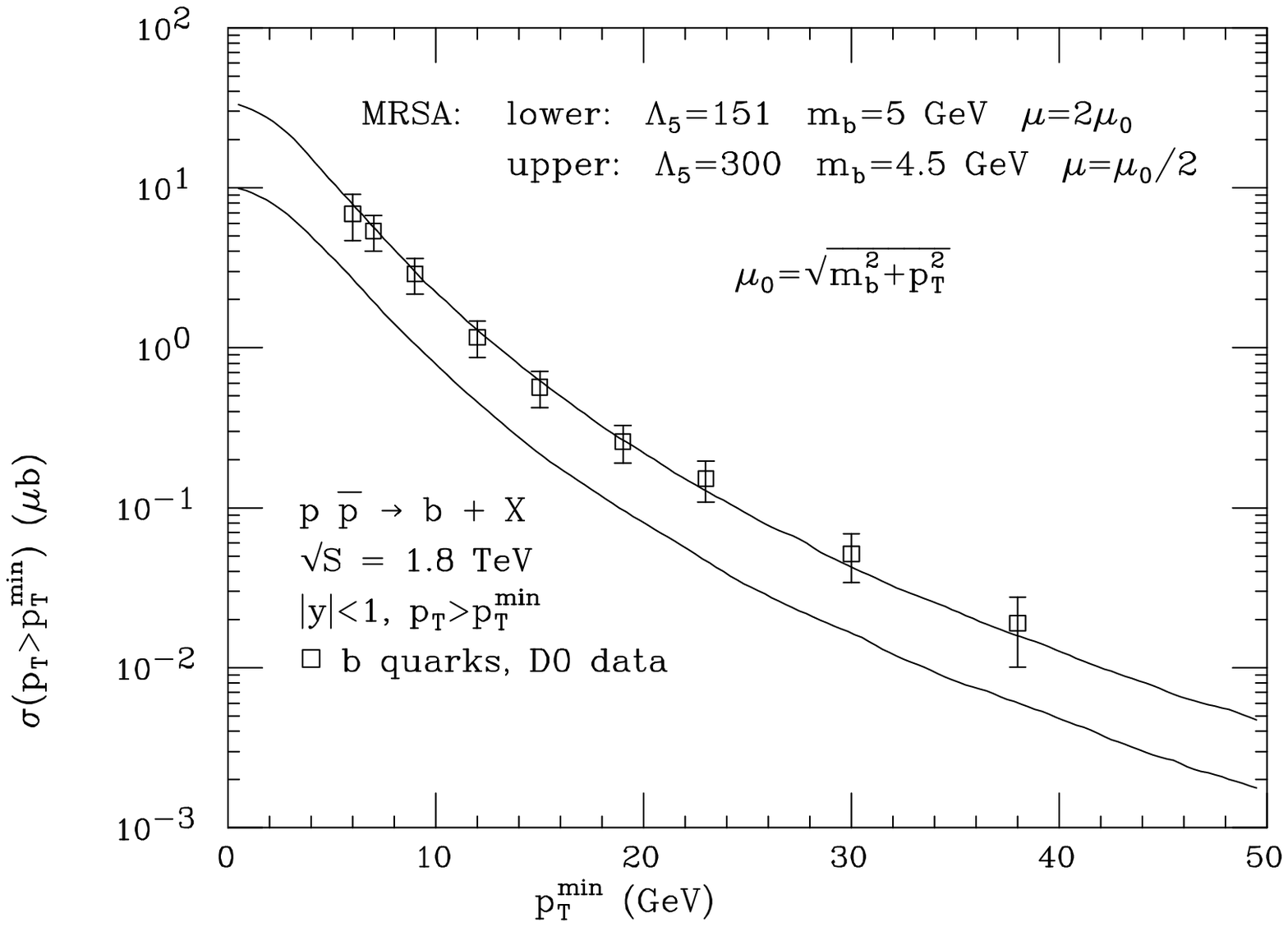,height=\bbigfig}}\end{center}
\caption{\label{d0bpt}
Cross section for $b$ quark production as a function of a
transverse momentum cut,
as measured by the D0 collaboration.
}
\end{figure}

\begin{figure}[htb]
\begin{center}\mbox{\epsfig{file=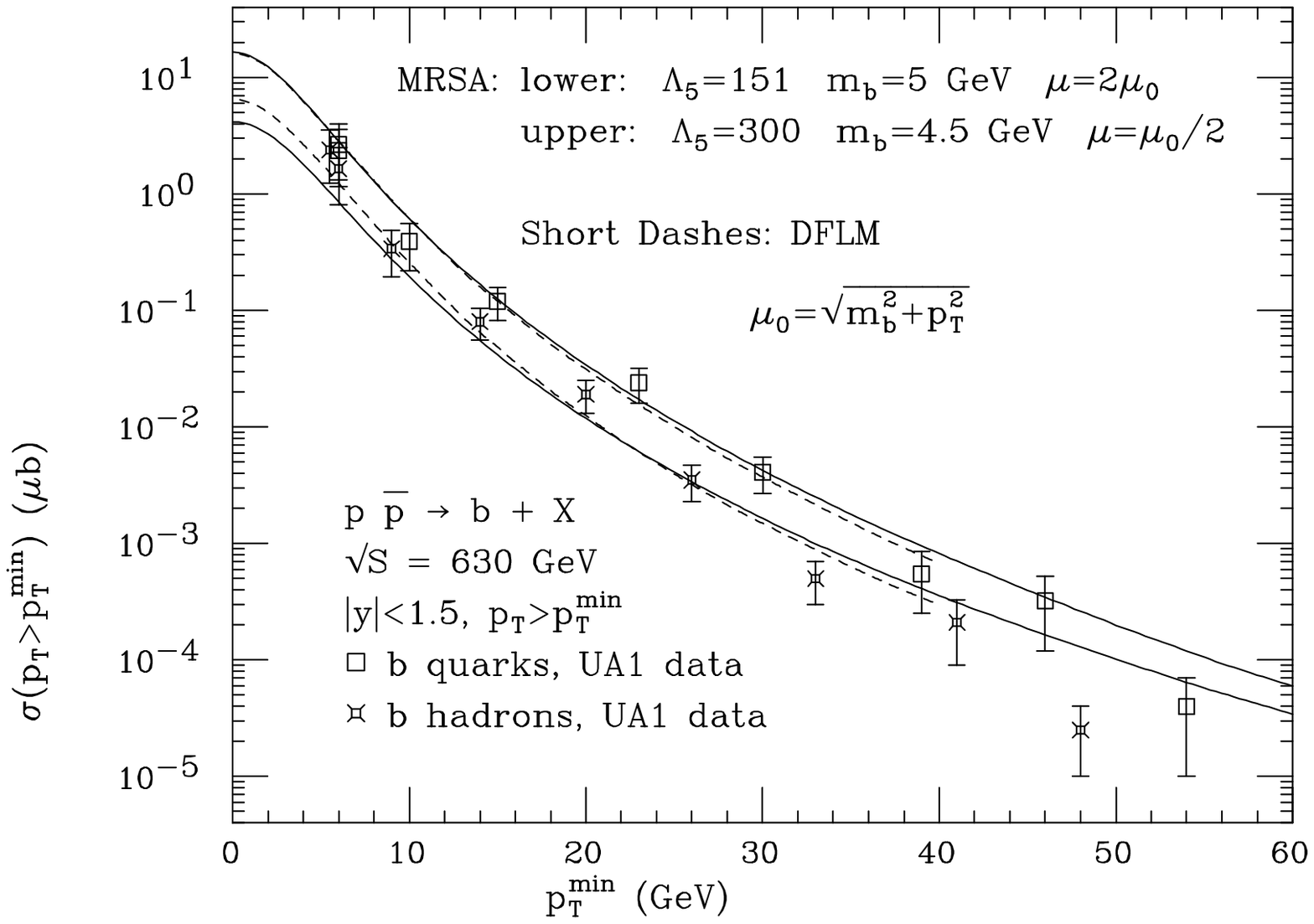,height=\bbigfig}}\end{center}
\caption{\label{ua1bpt}
Transverse momentum spectrum of B mesons at 630 GeV center of mass
energy, measured by the UA1 collaboration.
}
\end{figure}

In bottom production radiative corrections are large,
and various estimates of corrections
of even higher order (as given for example by the renormalization
and factorization scale dependence) lead to theoretical uncertainties
of the order of a factor of 2. These uncertainties, when
combined with other physical uncertainties, such as the error in the
knowledge of $\Lambda_{\rm QCD}$ and of the structure functions,
result in a rather poor theoretical prediction.
The CDF \cite{CDFbpt}, D0 \cite{D0bpt} and UA1 \cite{UA1bpt}
experiments have all measurements of the spectrum
of B mesons. The earliest CDF measurements reported a cross section
which was much higher than QCD prediction, and seemed to be in contrast
with CDF measurements. This problem was due to the poor theoretical
understanding of the direct $J/\Psi$ production, and after the introduction
of microvertexing techniques, the cross sections have come down to
smaller values.
The remaining discrepancy which is often quoted in CDF
publications is due to the fact that
modern sets of structure functions tend to favour small values for $\Lambda_5$,
and therefore smaller cross sections. If we instead allow
for larger values of $\Lambda_5$, as favoured by LEP experiments,
all we can conclude is that all data (including UA1 data) are
consistent with the theoretical prediction, although on the high
side of the theoretical band. What once seemed to be a
discrepancy between CDF and UA1 data, is now gone, both data sets bearing
the same relation with respect to the theoretical curves.
This is illustrated in fig.~\ref{cdfbpt}.
\settmpfig{NDE2}
 The main band (solid)
is obtained using the MRSA parton densities, except that for the
upper line the value of $\Lambda_5$ is taken to be 300 MeV, not quite
consistent with DIS data, but closer to what is indicated by LEP
measurements. The same curves are also shown without
fragmentation effects, in comparison with the full band (dotted lines)
obtained interpolating table 6 of ref.~\cite{NDE2}. As one can see
not much has changed since then in the theoretical prediction, and
the only thing one can say is that the CDF measurement is closer to
the upper limit of the theoretical band.
For comparison, I also show data from the D0 collaboration
in fig.~\ref{d0bpt}.
The old UA1 data is shown in fig.~\ref{ua1bpt}.
When comparing theoretical and experimental curves, one should
remember that certain experimental results are deconvoluted from the
heavy quark fragmentation effects, and are given as a bare quark cross
section, while others are heavy flavour meson's cross sections.
The former should be compared to the theoretical prediction obtained
without the inclusion of fragmentation effects, while for the latter
one should include fragmentation. The UA1 experiment
presented both a $B$ hadron cross section and a $b$ quark cross
section, by deconvoluting the effect of fragmentation.
In the figure
we show both the predicitons of ref.~\cite{NDE2}, and a more modern
one, performed using the MRSA structure functions, and the same variation
of parameters we used at Tevatron energy. Again, we see that the
theoretical band has not changed much. The UA1 quark data is on the
high side of the theoretical band, as the CDF and D0 data.
This fact has been quantified in the study of ref.~\cite{MLMPisa},
where it is found that the ratio of the data over theoretical upper band
is 0.97 for D0, 1.3 for CDF, and 0.84 for UA1.

We can conclude by saying that data on inclusive
$b$ production at hadron colliders
is consistent with QCD expectation, and lies on the upper
extreme of the prediction band. In other words,
in order to fit the cross section, one is forced to use values
of the parameters (like the scales, the mass, and the value of $\Lambda$)
that favour higher cross section.

As first pointed out in ref.~\cite{NDE}, the
perturbative computation of the $b$ cross section at hadron colliders
reaches a difficult kinematical regime when going from the
Sp$\overline{\mbox{p}}$S energies to the Tevatron. This is due to the
appearance of large logarithms of $S/m^2$ (where $S$ is the c.m. energy
squared, and $m$ is the mass of the heavy quark being produced) in the
perturbative expansion. In other words, at very large $S$ the
expansion parameter becomes $\as\log(S/m^2)$, which (for $S$ large
enough) becomes of order $1$, thereby spoiling the convergence
of the perturbative expansion.
Quite substantial theoretical work was performed
on the small-$x$ problem \cite{SmallxPapers}.
In particular, in ref.~\cite{CollinsEllis} it was found
that the resummation of small-$x$ effects increases the total rate
by 30\%. These effects are very likely to operate in the small $p_t$
region. It is unlikely, however, that they would strongly improve the
agreement between theory and data.

More detailed analysis of bottom distributions have been performed at the
Tevatron.
In refs.~\cite{CDFbazi} the azimuthal correlation has been measured and
compared with the theoretical prediction of ref.~\cite{MNR}.
While in qualitative agreement with QCD prediction, important differences
are observed. We can hope that correlation
measurements of this kind  may shed a light also on the problems
observed in the single inclusive $p_T$ distribution.

\section{Charm production}
\begin{figure}[htb]
\begin{center}\mbox{\epsfig{file=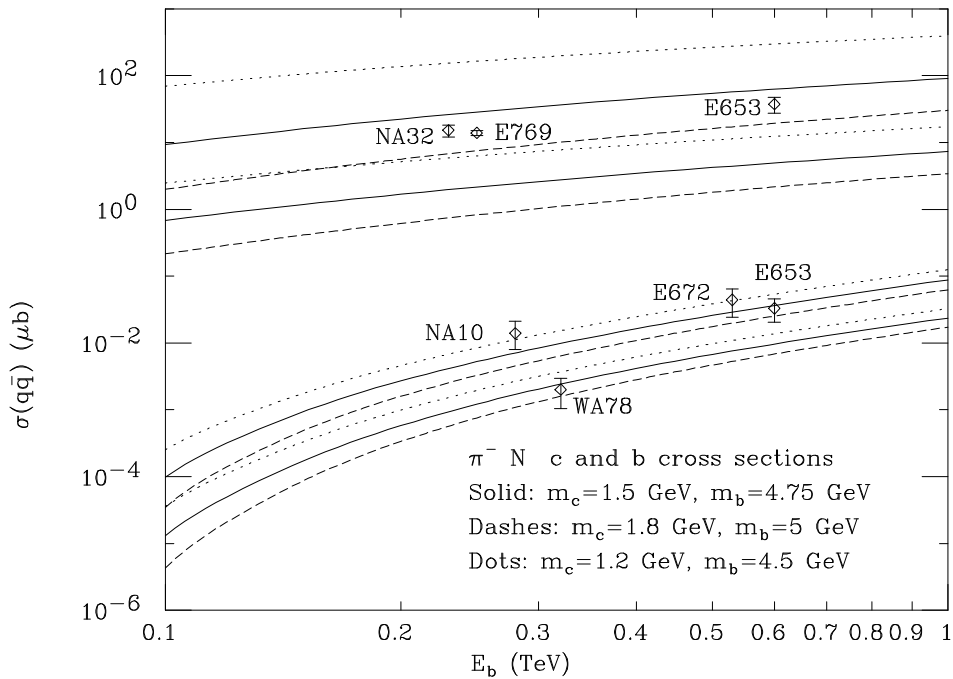,height=\bbigfig}}\end{center}
\caption{\label{bcpion}
Cross sections for $b$ and $c$ production in $\pi N$ collisions versus
experimental results.
}
\end{figure}

\begin{figure}[htb]
\begin{center}\mbox{\epsfig{file=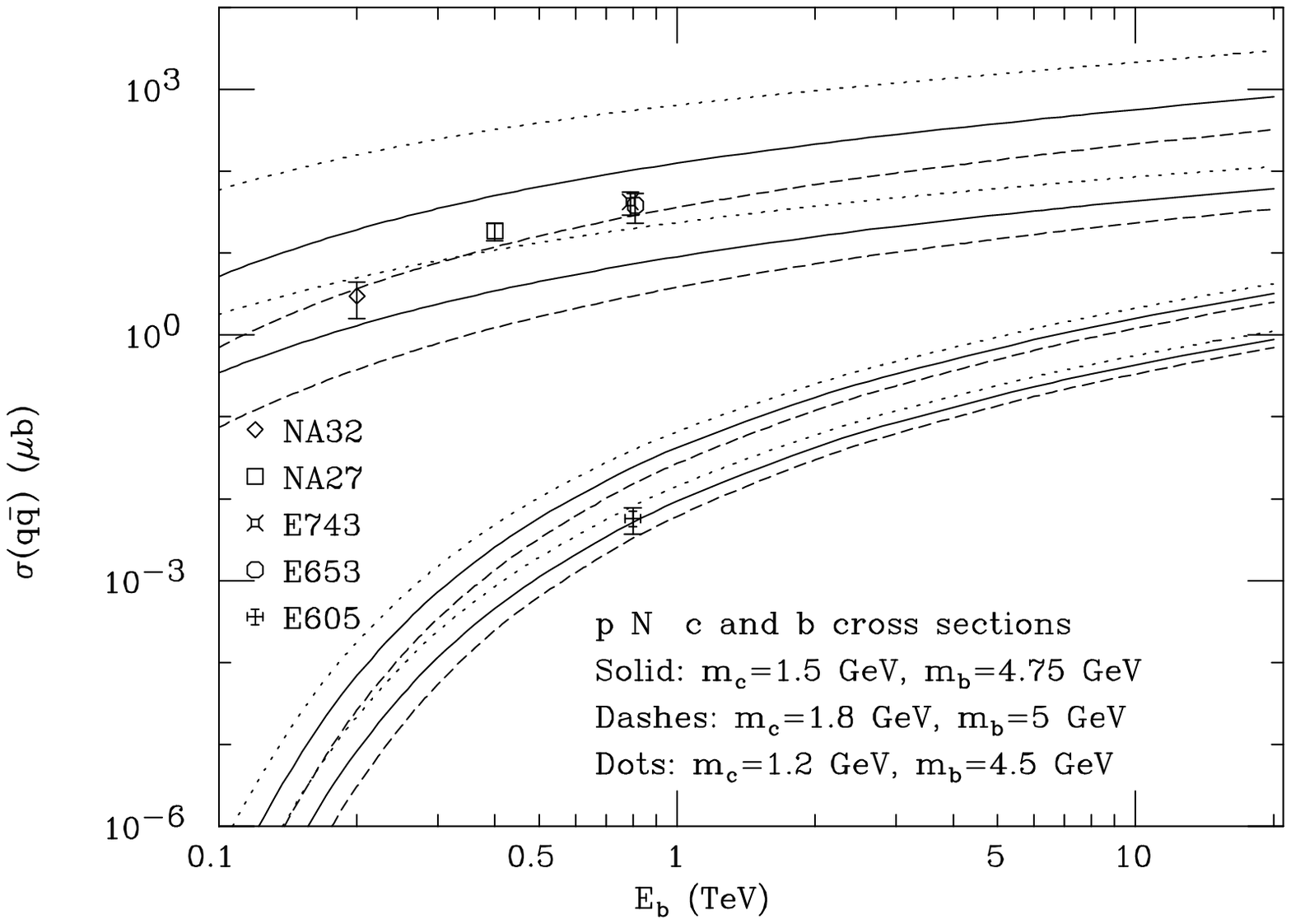,height=\bbigfig}}\end{center}
\caption{\label{bcproton}
Cross sections for $b$ and $c$ production in $p N$ collisions versus
experimental results.
}
\end{figure}

Figure~\ref{bcpion} gives an instructive picture of the uncertainties
in charm and bottom cross sections at fixed-target experiments.
Observe the considerable improvement that takes place when going from
charm to bottom. Observe also the strong mass dependence of the charm
result. Needless to say, similar uncertainties plague the $pN$ cross
sections, as shown in fig.~\ref{bcproton}.
The experimental measurement of bottom production cross section in
proton-nucleon collisions at fixed target is a new result \cite{E605}.
New-generation fixed-target experiments have accumulated very large statistics
of charm events. For a (possibly incomplete) list of recently published
results, see refs.~\cite{expincl} and \cite{expcorr}. For recent
reviews of the current experimental situation and future perspectives, see
refs.~\cite{appel}, \cite{introzzi}. For an older review see
\cite{Tavernier}. Results on bottom production at fixed target are
given in refs.~\cite{Albanese}--
\cite{Basile}.
As one can see,
experimental results on total cross sections for charm and bottom
production at fixed target are in reasonable agreement
with theoretical expectations.
We remind the reader that many puzzling ISR results in $pp$ collisions
at 62~GeV remain difficult to explain (see the review
\cite{Tavernier}),
in particular the large $\Lambda_b$ production rates reported
in ref.~\cite{Basile}.

\begin{figure}[htb]
\begin{center}\mbox{\epsfig{file=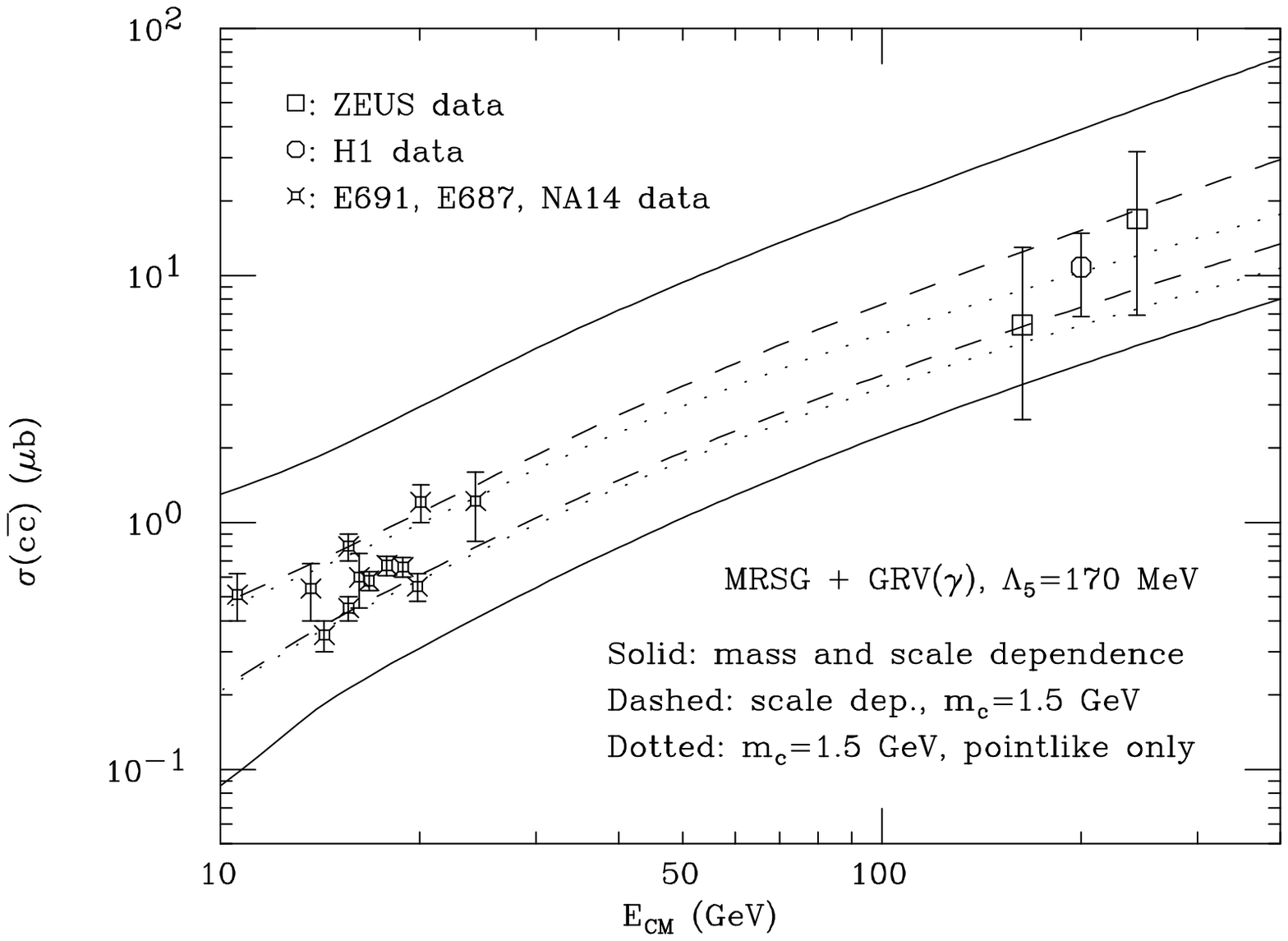,height=\bbigfig}}\end{center}
\caption{\label{phototot}
Cross sections for $c$ production in $\gamma N$ and $\gamma p$ collisions
versus experimental results.
}
\end{figure}

Photoproduction results are also in fairly good agreement with theory,
as can be seen from fig.~\ref{phototot}. The large band in the figure is
obtained by varying all parameters, including a variation of the charm
quark mass from 1.2 to 1.8 GeV. The dashed and dotted bands are instead
obtained by varying all other parameters, and keeping the mass of the
charm quark fixed to 1.5 GeV.
Very recently a new data point has been added by HERA
\cite{ZEUScharm}.
At the Bruxelles conference a new H1 result has been announced \cite{H1charm}.

\begin{figure}[htb]
\begin{center}\mbox{\epsfig{file=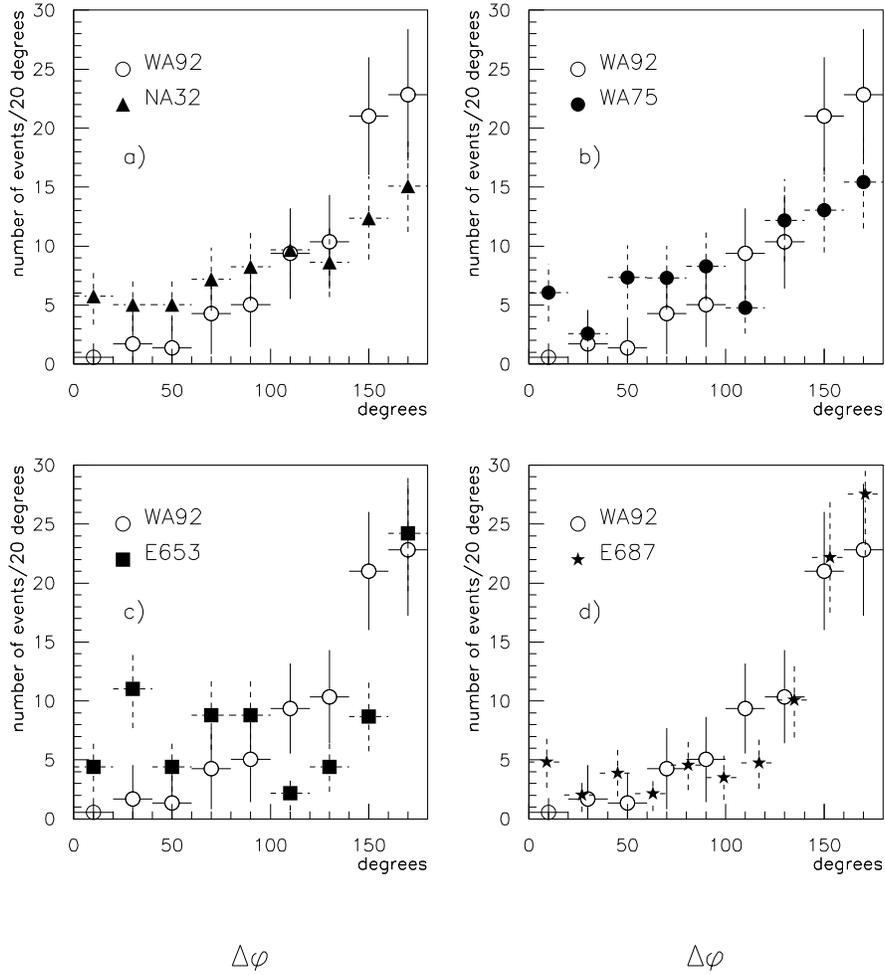,height=13cm}}\end{center}
\caption{\label{beatricefig7}
Azimuthal correlation in the production of charmed pairs.
}
\end{figure}

\begin{figure}[htb]
\begin{center}\mbox{\epsfig{file=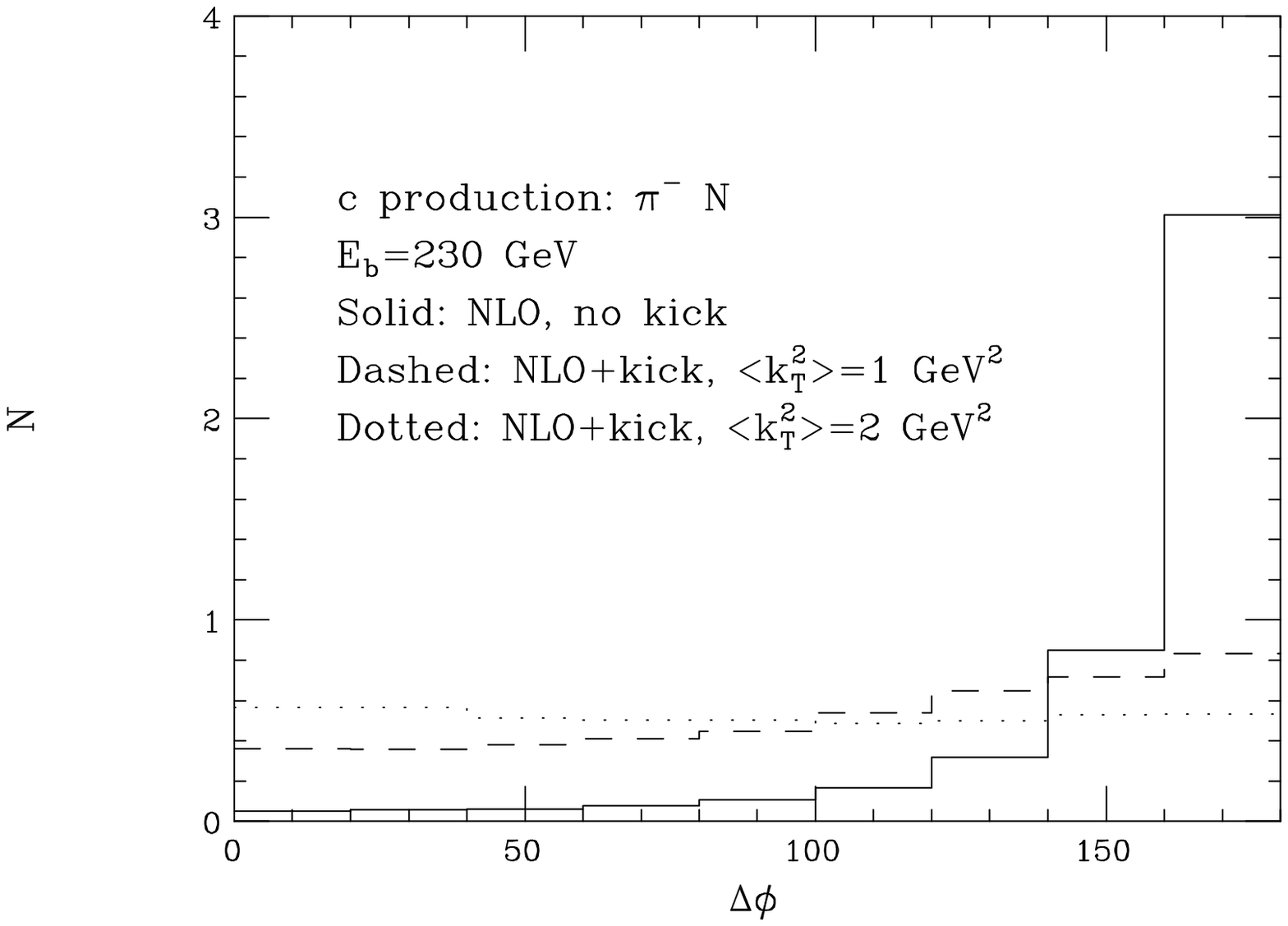,height=\bbigfig}}\end{center}
\caption{\label{azith}
Theoretical predictions for the azimuthal correlation
in the production of charmed pairs.
}
\end{figure}

Differential distributions for charm are at present in a more
complex situation. It turns out that in photoproduction one finds
a remarkably good agreement between theoretical expectations and
experiments. One typically looks at the transverse momentum
distribution of a single quark, the transverse momentum of the
pair, the invariant mass and the azimuthal correlation of the pair.
All this quantities are in good agreement with theoretical expectations,
provided one includes in the calculation the effect of a fragmentation
function, parametrized in the same way as in charm production in $e^+e^-$
collisions. This is not the case in hadroproduction. Typically one sees
that the inclusive $p_t$ distribution of a single meson is well described
by pure QCD, without the inclusion of a fragmentation function.
One also sees a similar behaviour for the $x_f$ distributions.
The azimuthal correlation of charmed pairs requires the inclusion
of some non-perturbative effects, that could be described as a primordial
transverse momentum of the incoming quarks. This transverse momentum
turns out to be of reasonable size, that is to say, below 1 GeV$^2$.
Another observable which is strongly sensitive to a primordial transverse
momentum of the incoming quarks is the transverse momentum distribution
of the charmed pair. In this case, only a very large primordial
transverse momentum could reproduce the measured cross section.
In the following I will illustrate this problems. It is fair to say that
a satisfactory answer to these problems is not known yet, but also that
present data gives some hints for a direction in which to deepen theoretical
work. I begin by showing in fig.~\ref{beatricefig7} the azimuthal distance
between the charm and anticharm, taken from ref.~\cite{beatrice}.
It is quite obvious that a strong enhancement in the back-to-back region
is observed, as expected from a hard production mechanism.
Perturbative calculations display a similar behaviour, although they are
generally much more peaked, as shown in fig.~\ref{azith}.
The solid histogram in the figure is obtain by a purely perturbative
calculation. The dashed and dotted histogram are obtained by assuming
that incoming partons do have some non-vanishing transverse momentum,
which is transmitted to the final quark--antiquark pair. This transverse
momentum can be viewed as a simple--minded model for non--perturbative
effects, due to the fact that hadrons have a finite transverse size.
The figure was obtained for a beam energy of 230 GeV, but for the sake
of the following discussion it can be considered as energy independent.
We see that all data seems to favour a $\langle k_T^2\rangle$ not larger
than 1 GeV$^2$, the WA92 data favouring even smaller values. A value
around 1 GeV$^2$ is roughly what one would expect, based upon experience
with Drell-Yan pair production, which requires intrinsic
transverse momenta of the order of 600 MeV$^2$.
Drell-Yan pairs are mainly produced
via quark--antiquark annihilation, while heavy flavours are mostly
produced by gluons. Since the gluons couple more strongly than quarks
by a factor of 9/4, one expects that the associated transverse momentum
should be somewhat larger. Observe that the azimuthal correlation is
independent of fragmentation effets, since fragmentation does not change
the direction of the outgoing particle. In fig.~\ref{beatricefig7},
data from the E687 photoproduction experiment is also reported \cite{E687etal}.
Comparisons of the E687 data with theoretical expectation are given in
ref.~\cite{FMNR}, and a good agreement is found, whether or not
one adds an intrinsic momentum kick to the incoming parton.
In fact, in the case of photoproduction, a transverse momentum kick has
a less dramatic effect, since only one parton is carrying the kick,
and also because the perturbative distribution is already broad, and it
washes out the effect of the kick.

\begin{figure}[htb]
\begin{center}\mbox{\epsfig{file=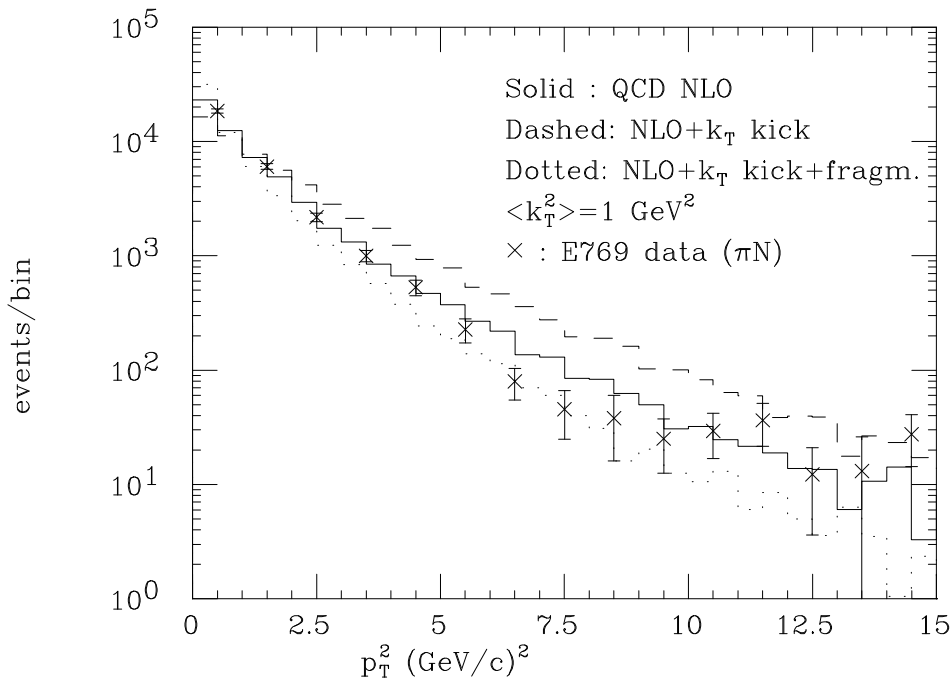,height=\bbigfig}}\end{center}
\caption{\label{e769pt2}
Single inclusive $p_T^2$ distribution measured by the E769 experiment,
compared to theoretical expectations.
}
\end{figure}

\begin{figure}[htb]
\begin{center}\mbox{\epsfig{file=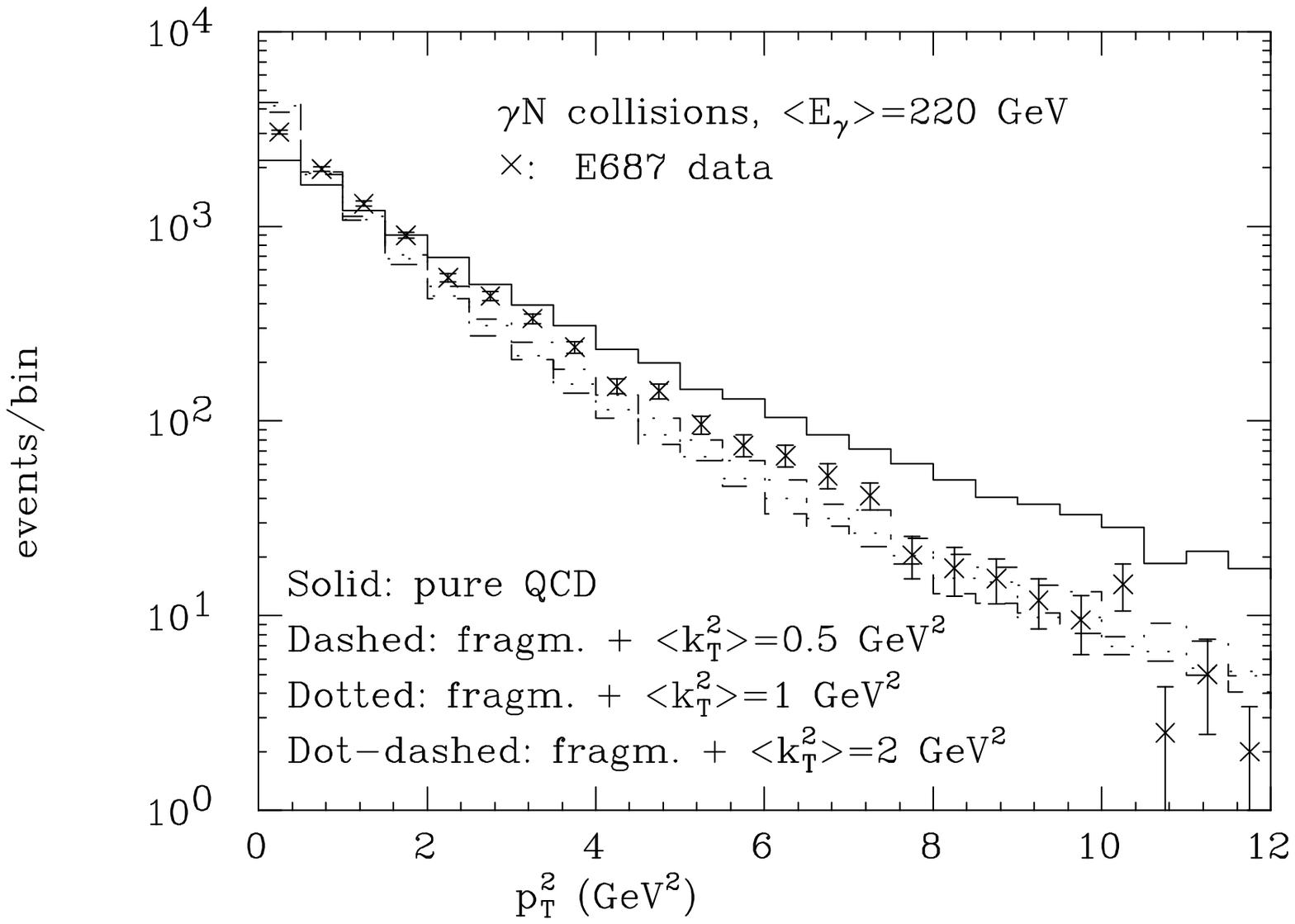,height=\bbigfig}}\end{center}
\caption{\label{e687pt2}
Single inclusive $p_T^2$ distribution measured by the E687 experiment,
compared to theoretical expectations.
}
\end{figure}

Let us now turn to the inclusive transverse momentum distributions.
Here we find a problem in the comparison of photoproduction and
hadropoduction data, as illustrated in figs.~\ref{e769pt2} and~\ref{e687pt2}.
We see that photoproduction data is in good agreement with QCD predictions,
including the effects of fragmentation. Transverse momentum
kicks do not alter the distribution in an appreciable way.
In the case of hadroproduction, one finds instead that the perturbative
prediction is in good agreement with data, and that the inclusion
of fragmentation effects spoil the agreement.
In the figure, also a transverse
momentum kick is included, which improves the agreement, but we still
see that even a 1 GeV$^2$ primordial transverse momentum is not enough
to overcome the effect of fragmentation.

A similar situation arises in the distributions of the transverse momentum
of the charmed pair. As an example I report in table~\ref{averages}
a computation of various average
quantities, in pion--nucleon collisions at 230 GeV, in comparison with
the recent study of ref.~\cite{Rybicki}.
These results deserve a comment. We see that, aside from the
rapidity difference, both the transverse momentum of the pair
and the invariant mass of the pair would be in better agreement
with the data if the effect of fragmentation was much less dramatic.
In fact, if we don't include fragmentation effects at all, we see that
all transverse distributions are in good agreement with theoretical
prediction, provided one is willing to accept the possibility of
a non--perturbative intrinsic transverse momentum effect of the incoming
parton of the order of 1 GeV$^2$, which is a
\begin{table}[htb]
\begin{center}
\begin{tabular}{|l||c|c|c|c|c|c|} \hline
& bare & \multicolumn{2}{c|}{$\langle k_T^2 \rangle=1\,{\rm GeV}^2$} &
         \multicolumn{2}{c|}{$\langle k_T^2 \rangle=2\,{\rm GeV}^2$} & data \\
&      & no fragm. & with fragm. & no fragm. & with fragm. &
 \\ \hline
$\langle p^2_{Q\bar{Q}} \rangle$
& 0.277 & 2.24 & 1.116 & 4.2 & 2.04 & $1.98\pm 0.11\pm 0.09$
 \\ \hline
$\langle M_{Q\bar{Q}}\rangle$
& 4.20 & 4.20 & 3.62 & 4.20 & 3.63 & $4.45\pm 0.03 \pm 0.13$
 \\ \hline
$\langle\Delta y_{Q\bar{Q}}\rangle$
& 0.931 & 0.886 & 0.669 & 0.852 & 0.650 & $0.54\pm 0.02 \pm 0.24$
 \\ \hline
\end{tabular}
\caption[htb]{  \label{averages}
Various average quantities computed with and without fragmentation
effects, and transverse momentum kick of 1 and 2 GeV$^2$, compared
with the results of the data analysis of ref.~\cite{Rybicki}.
}
\end{center}
\end{table}
very reasonable value.
It is often claimed that the fact that longitudinal distributions
(like the $x_F$ distributions) are harder than theoretical predictions
may require the inclusion of higher twist effects~\cite{Mueller}.
On the other hand, the problem is there only if one insists in using
the fragmentation functions for longitudinal distributions, a procedure
that is not fully justified. Many models are capable of justifying easily
the observed longitudinal distributions. On the contrary, the fact that
transverse momentum distributions are harder than expected is much
more worrying, since in this case the factorization theorem
should apply. One possible way out of this problem may be related to the
fact that a large fraction of bottom cross section in hadroproduction
comes from gluon fragmentation into heavy quarks, while in the case
of $e^+e^-$ annihilation, and also in the case of photoproduction,
this fraction is much smaller. This possibility requires further study,
but it could also explain why collider data for bottom production
is on the high side of the
prediction band. We have in fact seen that also in that case, if the
fragmentation function is not included, the agreement between theory and
data becomes much better.


\vskip 1cm
\begin{thebibliography}{99}
\bibitem{NDE}
  P.~Nason, S.~Dawson and R.~K.~Ellis, \np{B303}{88}{607}
\bibitem{NDE2}
  P.~Nason, S.~Dawson and R.~K.~Ellis,  \np{B327}{88}{49}.
\bibitem{Neerven}
  W.~Beenakker et al., Stony Brook preprint ITP-SB-90-46 (1990);
  \pr{D40}{89}{54}.
\bibitem{EllisNason}
  R.K.~Ellis and P.~Nason, \np{B312}{89}{551}.
\bibitem{SmithNeerven}
  J.~Smith and W.L.~Van Neerven, \np{B374}{92}{36}.
\bibitem{LaenenPh}
  E. Laenen, S. Riemersma, J. Smith and W.L. Van Neerven,
  \np{B392}{93}{162} and 229.
\bibitem{MNR}
  M.L. Mangano, P.~Nason and G.~Ridolfi, \np{B373}{92}{295}.
\bibitem{MNRFT}\label{FtMNR}
  M. Mangano, P.~Nason and G.~Ridolfi, \np{B405}{93}{507}.
\bibitem{FMNR}
  S. Frixione, M.~L.~Mangano, P.~Nason and G.~Ridolfi,
  \np{B412}{94}{225}.
\bibitem{CDFtop}
  F. Abe et al., CDF Coll., \prl{74}{95}{2626}.
\bibitem{D0top}
  S. Abachi et al., D0 Coll., \prl{74}{95}{2632}.
\bibitem{AltarelliTop}
  G. Altarelli, M. Diemoz, G. Martinelli and P. Nason,
  \np{B308}{88}{607}.
\bibitem{LaenenTop}
  E. Laenen, J. Smith and W.L. van Neerven, \np{B369}{92}{543};\newline
  \pl{321B}{94}{254}.
\bibitem{EllisTop}
  R.K. Ellis, \pl{B259}{91}{492}.
\bibitem{BergerTop}
  E.L. Berger and H. Contopanagos, preprint ANL-HEP-PR-95-31, hep-ph/9507363.
\bibitem{Schmelling}
  M. Schmelling, this proceedings (also CERN-PPE/95-129).
\bibitem{CDFbpt}
  F. Abe et al., CDF Coll.,
  preprint FERMILAB--Pub--95/48--E, March 18, 1995;\newline
  F. Abe et al., CDF Coll., \prl{68}{92}{3403}; {\bf 69}(1992)3704;
  {\bf 71}(1993)500, 2396 and 2537.
\bibitem{D0bpt}
  D0 Collaboration, S. Abachi et al., \prl{74}{95}{3548}.
\bibitem{UA1bpt}
  UA1 collaboration, \pl{B256}{91}{121}.
\bibitem{MLMPisa}
  M.L. Mangano, preprint CERN-TH/95-191, hep--ph/9508260,
  to appear in the Proceedings
  of the 6th International Symposium on Heavy Flavour Physics, Pisa,
  Italy, June 6-10, 1995.
\bibitem{SmallxPapers}
  R.K. Ellis and D.A. Ross, \np{B345}{90}{79};\newline
  S. Catani, M. Ciafaloni and F. Hautmann, \pl{B242}{90}{97},
  \np{B366}{91}{135}, {\it Nucl. Phys. B (Proc. Suppl.)}
  {\bf 23B}(1991)328.
\bibitem{CollinsEllis}
  J.C. Collins and R.K. Ellis, \np{B360}{91}{3}.
\bibitem{CDFbazi}
  F. Abe et al., CDF Coll., preprint FERMILAB--PUB--95/289--E,
  hep--ex--9508017;  \newline
  F. Abe et al., CDF Coll., preprint FERMILAB--PUB--94/131--E.
\bibitem{E605}
  D.M. Jansen et al., E605 Coll., \prl{74}{95}{3118}.
\bibitem{expincl}
  K. Kodama et al., E653 Coll., \pl{B284}{92}{461};\\
  S. Aoki et al., WA75 Coll., {\it Prog. Theor. Phys. }{\bf 87}(1992)1305;\\
  S. Barlag et al. NA32 Coll., \zp{C49}{91}{555};\\
  R. Ammar et al., E743 Coll., \prl{61}{88}{2185}.
\bibitem{expcorr}
  M. Aguilar-Benitez et al., NA27 Coll., \pl{164B}{85}{404};
  \zp{C40}{88}{321}; \\
  S. Aoki et al., WA75 Coll.,  \pl{209B}{88}{113};
  Prog. Theor. Phys. 87 (1992) 1315;\\
  K. Kodama et al., E653 Coll., \pl{263B}{91}{579}; CMU-HEP91-18(1991);\\
  S. Barlag et al., NA32 Coll., \pl{257B}{91}{519}.
\bibitem{appel}
  J.A. Appel, {\it Annu. Rev. Nucl. Part. Sci.} {\bf 42}(1992)367;\\
  L. Rossi, preprint INFN-AE-91-16, Oct. 1991,
  presented at 4th Int. Symp. on Heavy
  Flavour Physics, Orsay, France, 25-29 June 1991.
\bibitem{introzzi}
  G. Introzzi, E771 Coll., 3rd Topical Seminar on Heavy
  Flavours, San Miniato 17-21 June 1991, to appear in Nucl. Phys. B.
\bibitem{Tavernier}
  S.P.K. Tavernier, {\it Reports on Progress in Physics} {\bf
    50}(1987)1439.
\bibitem{Albanese}\label{pinbottom1}
  J. P. Albanese et al. (WA75 Coll.), \pl{108B}{82}{361}.
\bibitem{Basile}
  G. Bari et al., {\it Nuovo Cimento }{\bf 104A}(1991)1787;\newline
  M. Basile et al., {\it Nuovo Cimento }{\bf 65A}(1981)391.
\bibitem{ZEUScharm}
  The ZEUS Collaboration, preprint DESY 95--013, hep--ex--9502002.
\bibitem{H1charm}
  ``Photoproduction of
  $D^{*\pm}$ mesons in electron-proton collision at HERA'', H1 collaboration,
  1995 International Europhysics Conference on High Energy Physics,
  27/7 to 2/8/1995, Brussels, Belgium.
\bibitem{beatrice}
  The BEATRICE COLLABORATION, preprint CERN/PPE 94--214, DEC. 1994.
\bibitem{E687etal}
  M.P. Alvarez et al. (NA14/2 Coll.), \pl{B278}{92}{385},\\
  V. Arena et al. (E687 Coll.),   \pl{B308}{93}{194};\\
  J.C.~Anjos et al. (E691 Coll.), \prl{65}{90}{2503};\\
  M.I.~Adamovich et al. (Photon Emulsion Coll.), \pl{B187}{87}{437}
  and references therein.
\bibitem{Rybicki}
  K. Rybicki and R. Ry\l{}ko, Preprint BRU/PH/201, hep--ex--9505005.
\bibitem{Mueller}
  S.J. Brodsky, P. Hoyer, A.H. Mueller and W. Tang, \np{B369}{92}{519}.
\end{thebibliography}
\end{document}